# Local thermodynamic DOS measurement and twist-angle mapping in graphene-hBN superlattices

Namkyung Lee[1,2,a], Hangyeol Park[1,2,a], Seungwon Jung[1], Baeksan Jang[1,2], Seonyu Lee[1], and Joonho Jang[1,2,b]

[1] Department of Physics and Astronomy, Seoul National University, Seoul 08826, Korea

[2] Institute of Applied Physics, Seoul National University, Seoul 08826, Korea

[a] These authors contributed equally to this work.

[b] Author to whom correspondence should be addressed: joonho.jang@snu.ac.kr

**Abstract**

**Moiré patterns arising from twisted van der Waals stacks fundamentally reshape their electronic properties, enabling band-structure engineering that has driven rapidly growing interest in this field. In studying electronic properties, however, structural disorder present in real devices often leads to twist-angle inhomogeneity and obscures angle-dependent electronic effects when measured with bulk-averaged measurements. Probes that can access local thermodynamic response of the electronic systems with high sensitivity would be highly valuable. Here, we adopt Kelvin probe force microscopy (KPFM) to locally investigate graphene-hBN superlattices. By additionally modulating the chemical potential of the system, we obtain the inverse compressibility with high signal-to-noise ratio, enabling extraction of the local thermodynamic DOS. From this information, we determine the local twist angle along the device and find that twist-angle deviations are strongly correlated with bubble-induced strain features. Furthermore, by simultaneously tracking the offsets in the contact potential difference and in the net charge, we identify which interface within the heterostructure hosts the trapped bubbles. This capability to identify local electro-chemical environments provides a practical tool for strain-based studies and future device designs utilizing nanoscale engineering in moiré systems.**

When van der Waals (vdW) crystals are stacked to form a vdW heterostructure, the slight lattice mismatch or twist between adjacent layers gives rise to an emergent long-wavelength superlattice known as a moiré pattern.[1–3] This secondary periodic potential can drastically reshape the low-energy electronic band structure of the underlying system[4–6], generating electronic properties that are absent in the constituent materials. The ability to engineer the electronic band structure by the moiré potential has led to the realization of emergent correlated phenomena in vdW materials heterostructures, including superconductivity[7] and topological phases[8,9]. Development of practical applications based on the band-structure control has been proposed and partially demonstrated.[10–12] These advances further motivate systematic studies of the moiré systems under various conditions of twist angle, displacement field, and carrier density.

Despite the success in fabricating high-quality moiré systems, variation in their electronic behaviors of individual devices still remains undesirably high and sometimes the issues in the reproducibility of physics under investigation are reported.[13,14] This difficulty primarily arises from the metastable nature of the stacked configuration[15–17], which undergoes relaxation and in turn produces pronounced twist-angle inhomogeneity[18,19], obscuring the underlying electronic signatures. Bulk-averaged measurements, such as transport measurements, effectively average the signals from regions with different local twist angles, therefore smearing out the signatures originating from essential physics. From the viewpoint of device application, such inhomogeneity is a fundamental hurdle to construct and identify regions of the intended moiré patterns; profiling the local electronic environment and twist-angle landscape is crucial for fabricating devices with designed functionality. These issues motivate the use of scanning-based local probes that can access the electronic structure with spatial selectivity.

To address this challenge, in this letter, we implemented Kelvin probe force microscopy (KPFM) as the local-probe technique to directly access the local electronic properties, using a home-built setup that operates under cryogenic conditions.[20] Notably, KPFM probes the local electrical properties through a purely mechanical detection scheme.[21] In contrast to approaches such as single-electron transistors[22,23] or capacitance-based electrometry[24,25], where a metallic island or quantum dot is used as an electronic charge sensor, KPFM's mechanical detection simplifies the sensing element and inherently evades unnecessary electrostatic coupling to the device. Owing to this force-based detection principle, the measurement minimizes perturbation of the local electronic environment, thereby accessing genuinely thermodynamic properties in a non-invasive manner, while retaining spatial selectivity.

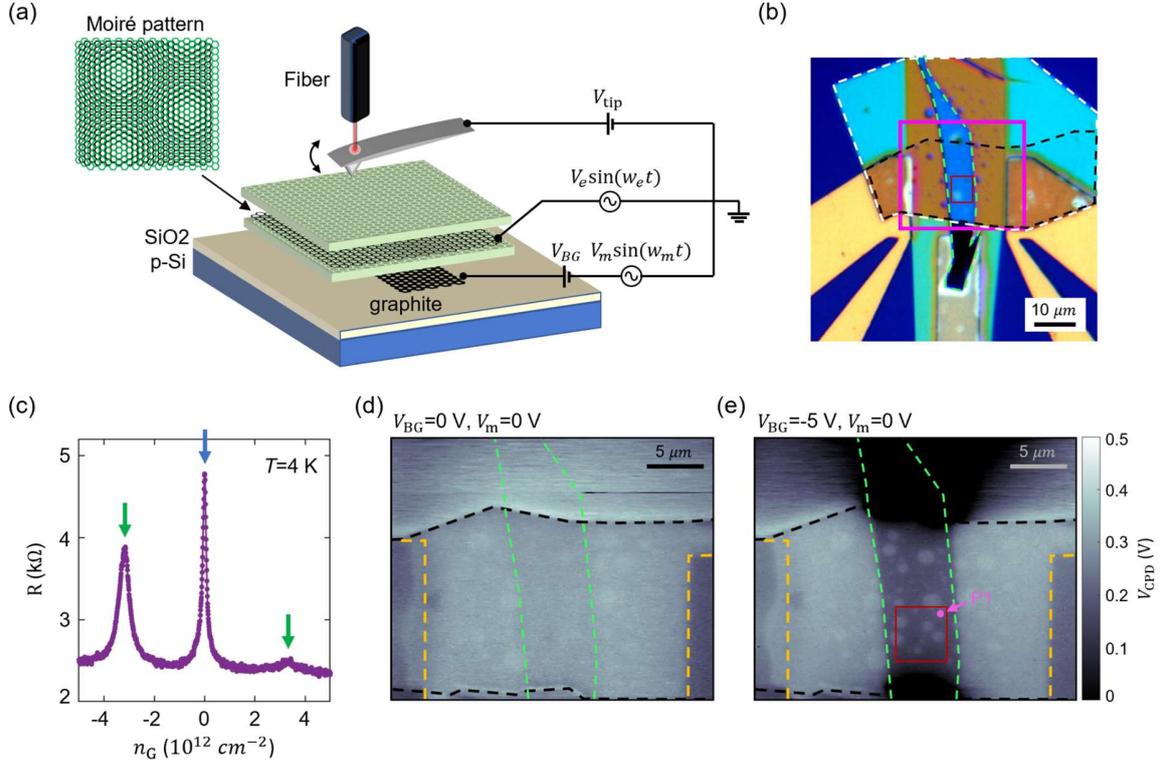

**Fig. 1** (a) Schematic of the KPFM setup on an hBN-encapsulated graphene stack (twist angle ~0.56°, top hBN aligned). (b) OM image of the device. Dotted outlines indicate the graphite back gate, graphene, and top hBN. (c) Two-probe resistance at 4 K versus carrier density, showing the charge-neutrality peak (blue arrow) and side peaks (green arrows) from the moiré potential in aligned graphene-hBN. (d, e) 2D KPFM maps from the region marked in panel (b), taken at $V_{BG}$=0 V and -5 V, respectively. Dashed lines indicate the same structural outlines as in panel (b). In panel (e), the red box marks the area used for the two-dimensional scan in Fig. 3, and the pink marker indicates the point used for the DOS trace in Fig. 2.

The measurement configuration is schematically illustrated in **Fig. 1(a)**. In this setup, DC bias and small AC voltages are applied to the tip, graphene layer, and the back gate, while the cantilever is monitored via laser interferometry. By identifying the voltage offset ($V_{\text{tip}}$) that nullifies the electrostatic force component at the modulation frequency ($f_e$), and thus the contact potential difference (CPD) between the tip and graphene, we obtain direct access to the chemical potential ($\mu$) of the system (see **Supplementary Materials** for details).[21] We then measure $\mu$ as a function of the back-gate bias voltage ($V_{BG}$) to extract $\mu$ versus carrier density $n$, allowing us to obtain the local electronic DOS of the moiré system. In our most sensitive measurements, we further superimpose an AC modulation ($V_m \sin(w_m t)$) onto the back-gate bias to directly access $d\mu/dn$ without a numerical derivative, which provides an additional enhancement in the signal-to-noise ratio (SNR).

Using this implementation of cryogenic KPFM, we scanned a hBN-encapsulated graphene-hBN heterostructure, as shown in **Fig. 1(b)**: The crystalline axes of top hBN (white dashed outline) and the graphene (black dashed outline) are aligned and stacked with a target angle of zero-degree, and they are placed on a graphite back gate (green dashed outline) underneath a bottom hBN layer. To characterize the bulk electrical response of the heterostructure, we performed a two-probe transport measurement, as shown in **Fig. 1(c)**. The trace at 4 K exhibits not only a primary resistance peak at charge neutrality (marked by blue arrow) but also additional side peaks on both the electron and hole sides (marked by green arrows), consistent with previous experimental reports[5,6] as well as with band-structure calculations for aligned graphene-hBN superlattices (see **Figure S4**). From the carrier densities required to fill the moiré minibands, the average twist angle of the device is estimated to be ~0.56°, a slight deviation from the target angle, potentially suggesting the spatial variation of the angle over the sample area.

To verify the spatial resolving capability of our measurement, we performed 2D scans over the pink solid box region of **Fig. 1(b)** at two fixed back-gate voltages, $V_{BG}$=0 V and -5 V, as shown in **Fig. 1(d)** and **1(e)**, respectively. In both maps, the black, green and orange dashed outlines indicate the regions of the graphene flake, the graphite back gate, and the Ti/Au electrodes, respectively. A clear change in the CPD distribution appears only over the graphene directly above the graphite gate when $V_{BG}$ is varied, while the rest of the graphene remains unaffected. This spatially selective response confirms that the measured signal faithfully tracks the back-gate-induced change of the graphene chemical potential ($\mu_G$)—rather than extrinsic environmental effects—thereby enabling quantitative extraction of a local electrical quantity with spatial resolution.

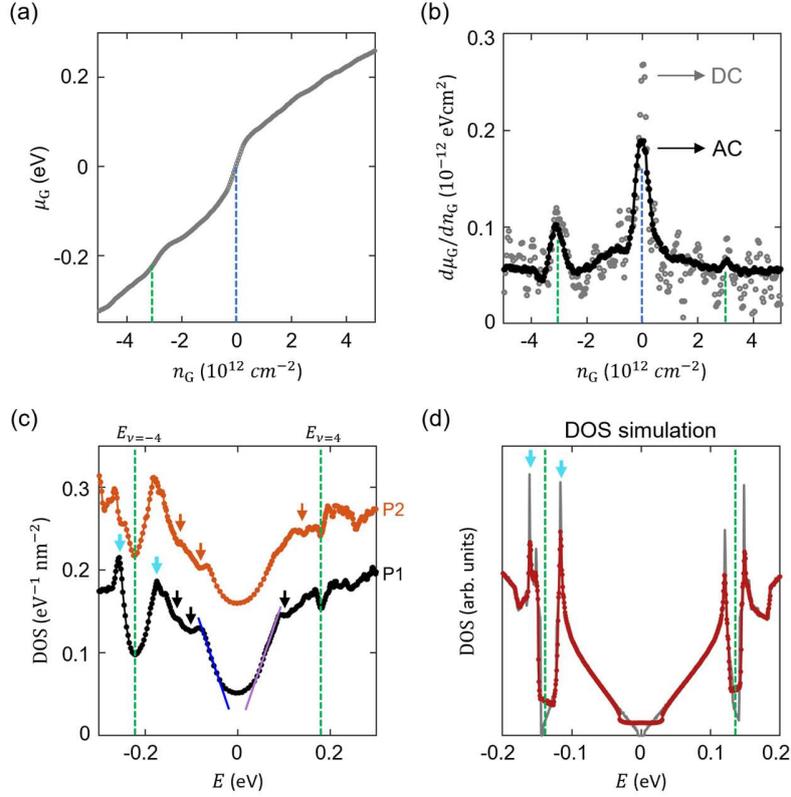

**Fig. 2** (a) Chemical potential $\mu_G$ as a function of carrier density $n_G$, measured at the P1 position indicated in Fig. 1(e), at 14 K. Vertical guides mark the CNP (blue) and the second Dirac point (SDP) feature (green). (b) Inverse compressibility $d\mu_G/dn_G$ obtained from numerical differentiation of (a) (grey) and from an AC-modulation measurement with a 0.5 $V_{RMS}$ drive on the back gate (black), the latter resolving the peak features with higher SNR. (c) DOS converted from the combined data in (a) and (b), taken at P1 (black) and at a nearby point P2 (orange; vertically shifted by 0.1 in the same units for clarity), showing DOS suppression at the same carrier densities where the peaks appear in (b). Cyan arrows mark vHS features near the hole-side SDP. (d) Continuum-model DOS (grey) and the DOS including a density-modulated broadening of one-tenth of the moiré full-filling density (red), both showing suppression at the CNP and at the SDPs, along with vHS peaks.

To obtain DOS, we measured $\mu_G$ as a function of graphene charge density $n_G$ at a fixed location (pink dot in **Fig. 1(e)**), as shown in **Fig. 2(a)**. In this process, we subtracted the background offsets in both charge density ($n_0$) and chemical potential ($\mu_0$)—originating from external environmental effects—by redefining $n_G = n - n_0$ and $\mu_G = \mu - \mu_0$ (see **Figure S2**). Consistent with the filling of electronic states, the $\mu_G$ increases monotonically with $n_G$, but the data exhibit pronounced slope changes as marked by blue and green dashed guides. When we plot the inverse compressibility, $d\mu_G/dn_G$, these features are more clearly identifiable. In **Fig. 2(b)**, we compare $d\mu_G/dn_G$ obtained by two different methods: one from numerically differentiating the data in **Fig.**

**2(a)** (grey dots), and the other from directly applying an AC modulation on the back gate (black). Comparing the two plots, the modulation-based measurement exhibits a substantially improved SNR—despite a slight peak broadening (see **Figure S3**)—making the relevant features more clearly discernable. In this representation, conspicuous peaks appear at CNP (blue dashed line) and at satellite positions symmetrically located about the CNP (green dashed lines), coinciding with carrier densities corresponding to full-filling of the moiré minibands. Since a low DOS value manifests as a maximum in inverse compressibility, the side peaks in $d\mu_G/dn_G$ indicate a DOS depletion associated with the formation of the moiré minibands.

One of the advantages of a sensitive KPFM is the ability to measure the accurate chemical potential difference. Under the assumption of a non-interacting picture that the chemical potential is identical to the single-particle energy[26], we extracted the local thermodynamic DOS as a function of energy ($E$) (from the same measurement set), in **Fig. 2(c)**. Two data sets with different spatial locations are presented; the black curve is measured at the very spot where the data in **Figs. 2(a, b)** were taken, whereas the orange curve is taken ~100 nm away. These data can be directly compared to the calculated DOS based on a continuum-limit model in **Fig. 2(d)** (see also **Supplementary Materials** for more details). The measured DOS shows qualitative agreement with the calculation, namely (i) suppressions at $E=0$ (CNP) and at the second Dirac points (SDPs), and (ii) pronounced van Hove singularity (vHS) peaks emerging in the vicinity of the SDPs (marked by cyan arrows).

Beyond the qualitative agreement, the electron-hole asymmetry in DOS is noticeable. For example, the energy locations of the SDPs on the electron and hole sides are clearly different. Furthermore, the Fermi velocities near CNP estimated from the measured data by the DOS-energy relation, as indicated in **Fig. 2(c)** (blue and purple solid lines), also show the asymmetry between the electron and hole dispersions. According to theoretical works on aligned graphene-hBN superlattices, the moiré potential renormalizes the Dirac dispersion of the graphene and reduce the Fermi velocity ($v_F$) near the CNP.[27–29] Relative to intrinsic graphene, the extracted slopes are larger by a factor of ~1.119 (electron side) and ~1.158 (hole side); using the relation $D(E) \propto |E|/(\hbar v_F)^2$, these correspond to $v_F$ reductions of ~5.5% and ~7.1%, respectively. These magnitudes of the suppression are comparable to values inferred from prior measurements.[25,30] The observed electron–hole asymmetry is likewise consistent with sublattice-asymmetric potentials, which in turn break $C_{2z}$ symmetry in aligned graphene–hBN superlattices.[31]

As the tip is moved to other positions, the overall shape of the DOS representing the unique dispersion of the moiré graphene[29], such as moiré-induced gaps, linear slope near CNP, and the vHS peaks at the hole band, remains. On the other hand, there are noticeable features that are not easily accounted for by the ideal band dispersion; weak DOS suppressions at distinct energies are clearly visible (marked by black and orange arrows in **Fig. 2(c)**), and their shapes and energetic locations change significantly depending on the measurement position. We interpret these features in DOS as a consequence of local charge impurities in our device[32]. Overall, the DOS measured at various spots provides valuable information on the spatial inhomogeneity of this sample and its impact on the electronic spectra of this moiré graphene system.

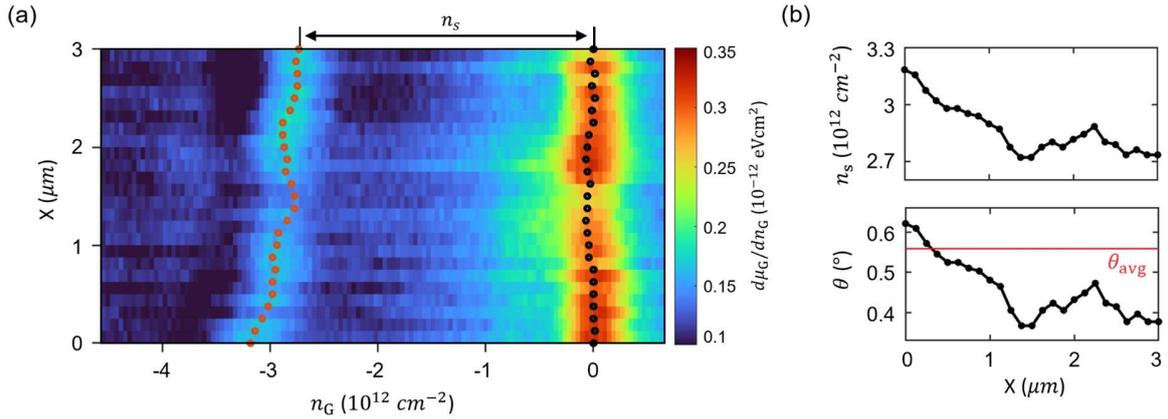

**Fig. 3** (a) Spatial evolution of $d\mu_G/dn_G$ as a function of $n_G$ and tip position, showing continuous shifts of the CNP peak and the moiré side peak. (b) (top) Density separation $n_s$ between the CNP and the side peak extracted from (a), plotted versus position. (bottom) Local twist angle inferred from $n_s$, with the red line indicating the average twist angle obtained from transport.

Through the high sensitivity of our KPFM to subtle DOS variations, the analysis can be extended beyond a single-point measurement (see **Fig. 2**) to a spatially resolved line scan. In **Fig. 3(a)**, the inverse compressibility ($d\mu_G/dn_G$) as a function of $n_G$ is mapped along a one-dimensional trajectory across the graphene region—taken along the lower edge of the red-boxed area in **Fig. 1(e)**. The data reveal two pronounced local maxima, corresponding to the CNP and the hole-side SDP[29], respectively. The separation between them in carrier density ($n_s$) varies continuously along the scan, signaling the spatial inhomogeneity of the twist angle across the sample. The extracted evolution of $n_s$ is plotted in the top panel of **Fig. 3(b)**. Since $n_s$ quantifies the carrier filling per moiré unit cell, its spatial variation directly reflects a change in the moiré length, $l_m = a/\sqrt{\delta^2 + \theta^2}$, which arises from a variation in the local twist angle. The measured $n_s$ can therefore be converted into a position-dependent twist-angle profile, as shown in the bottom panel of **Fig. 3(b)**, distributed around the twist angle extracted from transport ($\theta_{\text{avg}}$, marked by a red solid line) and revealing local information that is unavailable in transport measurements.

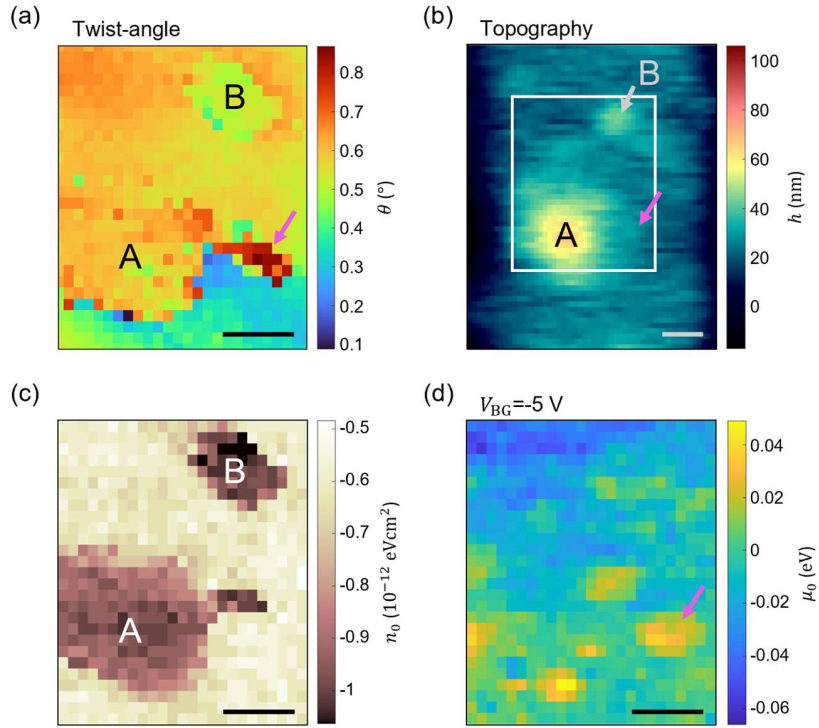

**Fig. 4** (a) 2D twist-angle map showing visibly distinct twist-angle regions (A and B), together with several sharply varying points, one of which is marked by the pink arrow. (b) Topography over a larger area, with the white solid box marking the region corresponding to the scan in (a). Large bubbles are located at A and B, and a smaller bubble is seen at the pink arrow location. (c, d) Concurrent 2D maps of the charge-density offset $n_0$ (c) and the chemical-potential offset $\mu_0$ measured at $V_{\text{BG}}=-5$ V (d), taken over the same area as (a). In (c), the dominant features coincide with the large bubbles at A and B, whereas in (d), no such features appear at the locations of A or B, and instead a feature is observed at the pink arrow location. All scale bars, 1 μm.

We extend the twist-angle measurement to a two-dimensional scan over the region marked by the red box in **Fig. 1(e)**, as shown in **Fig. 4(a)**. In this map, two regions labelled A and B exhibit visibly distinct twist angles. The corresponding topography shown in **Fig. 4(b)** indicates that bubbles are trapped at the locations of A and B, suggesting that bubble-induced local strain gives rise to the twist-angle variation. In particular, the twist angle is increased at A whereas it is reduced at B relative to the surrounding region, implying that bubble-induced strain can locally shift the twist angle in either direction. Notably, apart from regions A and B, the twist-angle map reveals several sharply varying points, one of which is marked by a pink arrow in **Fig. 4(a)**. In the topography, a bubble is likewise observed at the location (pink arrow in **Fig. 4(b)**), albeit smaller than those at A and B. This small bubble affects the twist angle of the graphene much stronger than the larger bubbles at A and B do. This observation thus suggests that the smaller bubble lies much closer to the graphene, highly likely at the graphene interfaces—whereas the larger bubbles at A and B may reside at a deeper interface such as the graphite-hBN interface, thereby exerting a weaker effect on the twist angle.

To further identify at which interface the bubbles reside, we measured the spatial distribution of the offsets in chemical potential and charge density ($\mu_0$ and $n_0$, respectively; see also **Figure S2**). These quantities, in principle, indicate the variations in the local electrical environment due to the external sources to the system, such as charge puddles, hydrocarbon contaminants, and interfacial bubbles. A shift of $n_0$ reflects a modulation of the local charge density influenced by various types of charge reservoirs and bubbles present at any interface[33,34]. In contrast, a shift of $\mu_0$ most likely indicates a contact potential difference induced by foreign species in direct contact with the graphene[35,36], in this case, likely bubbles residing at interfaces adjacent to graphene. **Figs. 4(c, d)** show the spatial maps of $n_0$ and $\mu_0$, respectively, for the same area. In **Fig. 4(c)**, the dominant features in $n_0$ are strongly correlated with the large bubbles visible in the topography (marked by A and B in **Fig. 4(b)**), indicating that the large bubbles host significant trapped charges that can influence the carrier density in the nearby graphene system.

By contrast, the map of $\mu_0$ in **Fig. 4(d)** does not show any features related to the large bubbles; instead, it displays distinct features associated with smaller bubbles. This observation strengthens the idea that only the small bubbles have direct contact with the graphene layer—probably at the bottom hBN-graphene interface, as implied by the presence of a moiré band feature—whereas the large bubbles influence the graphene only through the remote charge effect. Thus, we conclude that the small bubbles are at the interface between the bottom hBN and the graphene, while the large bubbles indicated by A and B are located between the graphite back gate and the bottom hBN. Therefore, our technique and the analysis based on the chemical potential and charge offsets provide invaluable information on the graphene heterostructure, combined with the local thermodynamic DOS and twist-angle estimations.

In conclusion, we measured local thermodynamic DOS using a high-sensitivity cryogenic KPFM. The data demonstrate that the electronic DOS serves as a reliable indicator of spatial twist-angle inhomogeneity in a graphene-hBN moiré device, enabling the reconstruction of its twist-angle landscape. Moreover, the twist-angle modulation is notably enhanced at bubble sites, and the offset measurements concurrently reveal the vertical interface at which those bubbles reside. This combined capability offers device-relevant insight into strain-induced twist-angle disorder and provides a practical route for future strain-resolved studies in moiré systems.

## Supplementary material

See the supplementary material for experimental details and supporting figures.

## Acknowledgments

This work was supported by the National Research Foundation of Korea grants funded by the Ministry of Science and ICT (Grant Nos. RS-2025-23525425, RS-2020-NR049536, and RS-2023-00258359), SNU Core Center for Physical Property Measurements at Extreme Physical Conditions (Grant No. 2021R1A6C101B418), and Creative-Pioneering Researcher Program through Seoul National University. K.W. and T.T. acknowledge support from the JSPS KAKENHI (Grant Numbers 21H05233 and 23H02052) and World Premier International Research Center Initiative (WPI), MEXT, Japan.

## Author declarations

### Conflict of interest

## Data availability

The data that support the findings of this study are available from the corresponding authors upon reasonable request.

# Supplementary material for "Local thermodynamic DOS measurement and twist-angle mapping in graphene-hBN superlattices"

Namkyung Lee[1,2,a], Hangyeol Park[1,2,a], Seungwon Jung[1], Baeksan Jang[1,2], Seonyu Lee[1], Joonho Jang[1,2,b]

[1] Department of Physics and Astronomy, Seoul National University, Seoul 08826, Korea

[2] Institute of Applied Physics, Seoul National University, Seoul 08826, Korea

[a] These authors contributed equally to this work.

[b] Author to whom correspondence should be addressed: joonho.jang@snu.ac.kr

## I. Device fabrication

The device was fabricated using a layer-by-layer assembly technique. The process consisted of the following five steps. (1) A graphite flake was mechanically exfoliated and transferred onto a pre-patterned Ti/Au (5 nm/70 nm) electrode using a polydimethylsiloxane (PDMS) stamp. (2) An exfoliated hBN flake was then transferred onto the graphite using the same PDMS-based method, yielding an hBN/graphite stack. (3) Using a PDMS/polycarbonate (PC) stamp, an hBN flake and subsequently a graphene flake was sequentially picked up. During this step, the crystallographic axes of graphene and hBN were carefully aligned. The resulting stack was then released onto the pre-assembled hBN/graphite using the glass-transition-assisted release of PC. (4) The residual PC was removed by soaking in N-methyl-2-pyrrolidone (NMP), followed by a rinsing with isopropyl alcohol (IPA). (5) Finally, to improve the interface quality and promote further alignment relaxation, the assembled device was annealed at 300 °C for ~ 10 minutes.

## II. KPFM principles and data processing

### II.1. Electrostatic force nulling in KPFM

When a combined bias $V = V_{DC} + V_{AC}\sin(w_e t)$ is applied between the tip and the graphene, the cantilever experiences an electrostatic force $F = \frac{1}{2}\frac{\partial C}{\partial z}(V - V_{CPD})^2$, where $V_{CPD}$ is the contact potential difference between the tip and the sample. This force can be decomposed into a static term and oscillatory terms at $\omega_e$ and $2\omega_e$. The first-harmonic component is $F_w = \frac{\partial C}{\partial z}(V_{DC} - V_{CPD})V_{AC}\sin(w_e t)$, containing the information on the contact potential difference. By tuning $V_{DC}$ to null this term, one obtains $V_{CPD}$.

In this study, the nulling procedure was repeated while sweeping the back gate. Since the back gate controls the carrier density ($n$) of the graphene, the resulting variation of $V_{CPD}$ reflects the corresponding change in the chemical potential ($\mu$).

### II.2. Gate-voltage-to-density conversion

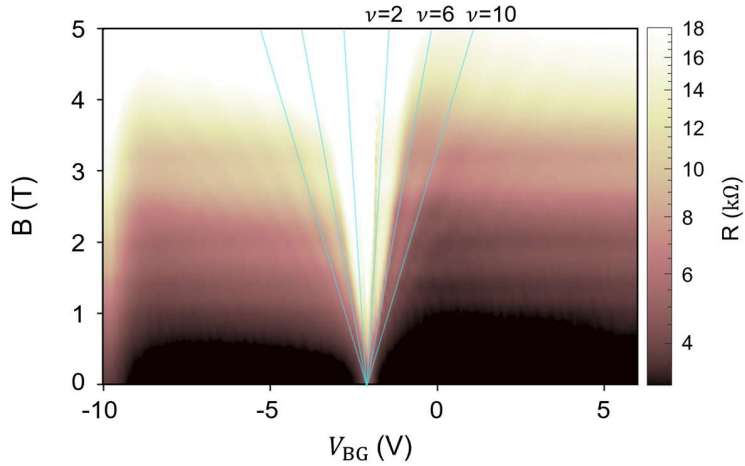

**Figure S1** Resistance measured as a function of perpendicular magnetic field and back-gate voltage. Landau fans originating from the CNP appear at filling factors of ±2, ±6 and ±10, where the resistance drops along the fan trajectories. Cyan guides mark the fan position corresponding to these filling factors.

To quantitatively convert the back-gate voltage into carrier density, we performed magneto-transport measurements at 4 K. The two-probe resistance as a function of magnetic field and back gate voltage is shown in **Figure S1**. A pronounced resistance maximum appears near $V_{BG}$=-2 V, identifying the CNP. From this point, Landau fans with filling factors 2, 6, 10, … emerge from the CNP. By reading off the fan slopes in the $(B, V_{BG})$, we determined the gate-voltage-to-density conversion factor $dn/dV_{BG}$ to be 0.408 ($10^{12}$ cm$^{-2}$V$^{-1}$), which is comparable to the value inferred from the geometric capacitance.

## II.3. Offset removal and extraction of $n_G$ and $\mu_G$

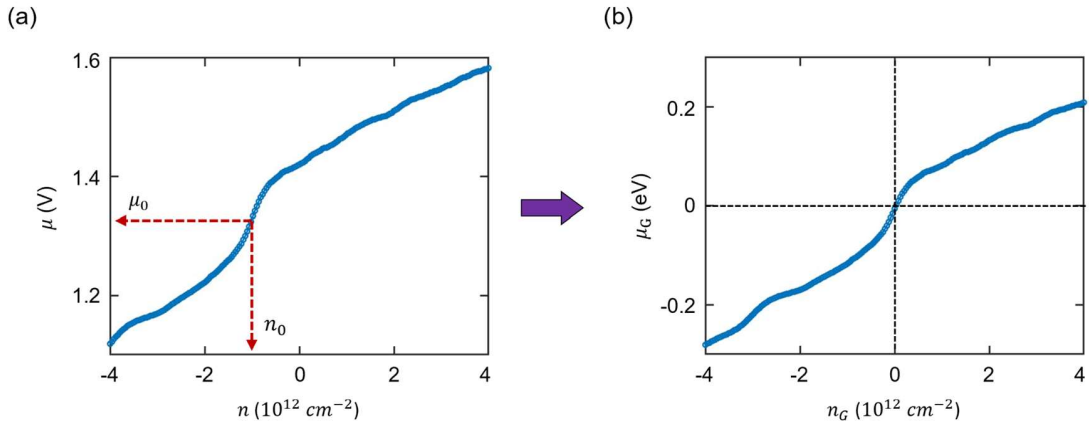

**Figure S2** Schematic illustrations of the offset-removal procedure used to extract $\mu_G$ and $n_G$. (a) Raw $\mu$ versus $n$ before offset removal. (b) $\mu_G$ versus $n_G$ after the correction is applied.

Local electrostatic environments can introduce finite offsets in the measured carrier density and chemical potential. To recover the intrinsic $\mu$-$n$ relation, we removed these offsets prior to analysis. **Figure S2(a)** shows the raw $\mu$-$n$ plot, where $\mu$ is taken from the measured $V_{CPD}$ and $n$ is obtained from $C_g V_{BG}/e$. A sharp change in slope appears near $n$=-1 ($10^{12}$cm$^{-2}$), identifying the CNP. We take the values of $\mu$ and $n$ at this point as $\mu_0$ and $n_0$. After subtracting these offsets, i.e. defining $n_G = n - n_0$ and $\mu_G = \mu - \mu_0$, the resulting offset-corrected $\mu_G$-$n_G$ plot is shown in **Figure S2(b)**. This referencing ensures that all quantities are aligned to the CNP and are not affected by device-dependent electrostatic shifts.

## II.4. Broadening of DOS induced by AC modulation

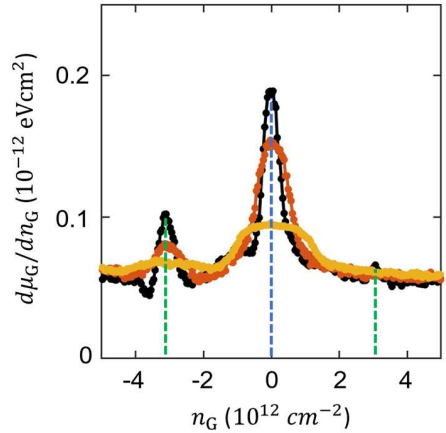

**Figure S3** Inverse compressibility trances measured with an AC modulation applied to the back gate. Black, orange, and yellow curves correspond to modulation amplitudes of 0.5 V, 1 V, and 2 V, respectively. Increasing the modulation amplitude leads to broader peak features at the CNP (blue dashed line) and second Dirac points (green dashed lines).

When an AC modulation is applied to the back gate to obtain inverse compressibility $d\mu_G/dn_G$, the carrier density oscillates within a finite window proportional to the modulation amplitude. This produces an effective averaging along the density axis and can broaden sharp features in the inverse compressibility (and consequently in the DOS).

To quantify this effect, we measured $d\mu_G/dn_G$ while varying the modulation amplitude (**Figure S3**). The black, orange, and yellow curves correspond to AC amplitudes of 0.5 $V_{RMS}$, 1 $V_{RMS}$, 2 $V_{RMS}$, respectively, which translate to carrier-density windows of approximately 0.58, 1.15, and 1.73 ($10^{12}$ cm$^{-2}$). As the amplitude increases, the spectral peaks become noticeably broader.

Thus, while AC modulation suppresses low-frequency noise and improves sensitivity, it also introduces a trade-off in spectral resolution due to broadening. Our setup can therefore be operated either in an AC mode (sensitivity-oriented) or a DC mode (resolution-oriented), depending on the experimental objective.

## III. Band structure simulation

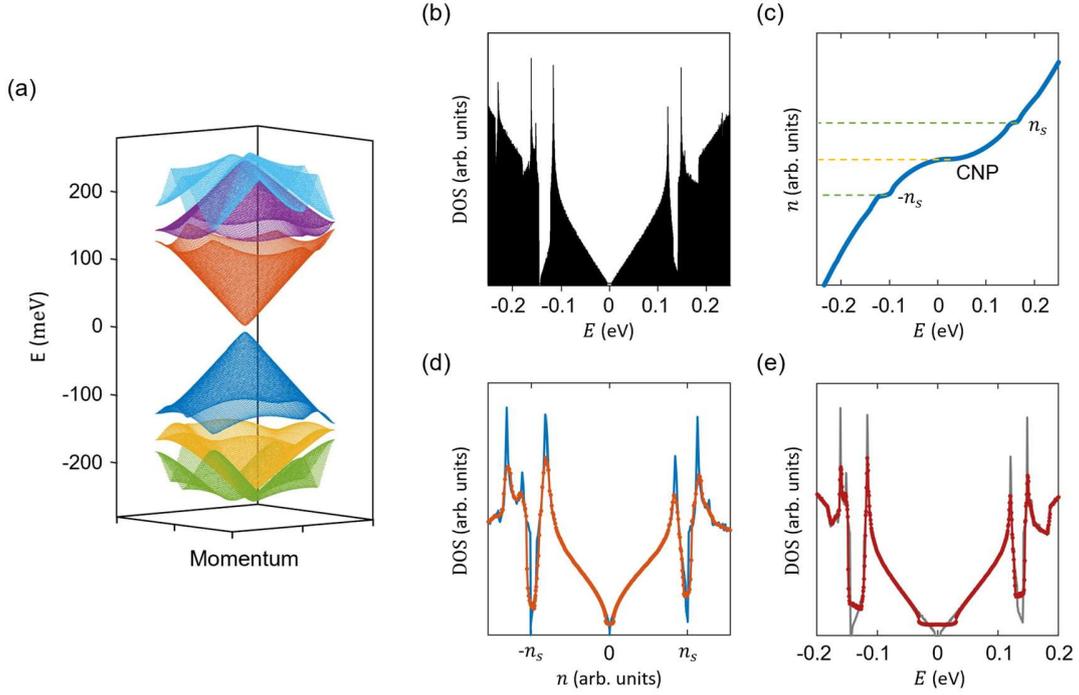

**Figure S4** (a) Continuum-model band structure near the CNP, in the moiré Brillouin zone, shown as a 3D surface. (b) DOS as a function of $E$ obtained from the same calculation. (c) $n$ plotted as a function of $E$ converted from (b). (d) DOS versus $n$ derived from the model (blue), together with a smoothed trace obtained by averaging over a finite density window (one-tenth of full-filling). (e) DOS versus $E$ shown for comparison. The grey curve is the direct output of the calculation, whereas the red curve is obtained by applying the same finite-density-window averaging. This panel corresponds to Fig, 2(d) in the main text.

We performed band structure calculations to provide a theoretical reference for the DOS analysis, following the procedure used in a previous work from our group[37] and standard continuum-based approaches in the literature[38]. The Hamiltonian was diagonalized over the moiré Brillouin zone using the continuum model introduced by Bistritzer and MacDonald. The resulting band structure near the CNP is shown in **Figure S4(a)**, exhibiting a gap at the CNP and at the hole-side second Dirac point (SDP). From this dispersion, we computed the DOS as a function of energy (**Figure S4(b)**) and subsequently obtained *n(E)* by integrating DOS over energy (**Figure S4(c)**). Using these quantities, we plot DOS versus *n* and DOS versus *E* in **Figures S4(d, e)**. In these plots, the blue/grey curves represent the direct output of the model, while the orange/red curves incorporate finite-density-window averaging (one-tenth of full filling) to emulate the experimental broadening. The simulated trends are qualitatively consistent with previous continuum-model results reported in the literature.[27-29]

## IV. Interpretation of the offsets $n_0$ and $\mu_0$

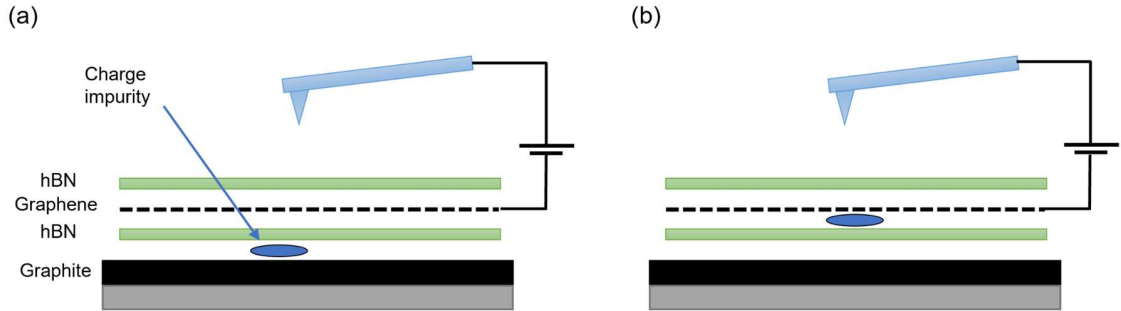

**Figure S5** Schematic illustration of two representative impurity configurations: (a) a charge impurity trapped at an interface away from the graphene (e.g. between the graphite and the bottom hBN), and (b) a charge impurity in direct contact with the graphene.

Local electrostatic environments can shift the measured carrier density and chemical potential, giving rise to the offsets $n_0$ and $\mu_0$. The effect of charge impurities depends sensitively on where they are located along the vertical stack. Two representative configurations are illustrated in **Figure S5**.

In the first case (**Figure S5(a)**), a charged impurity is trapped at an interface away from the graphene (for example, between the graphite and the bottom hBN), so that it does not make electrical contact with the graphene. Such a charge primarily acts as an external dopant and is therefore expected to shift $n_0$. However, because such an impurity is located away from the tip-graphene system, it does not affect the contact potential difference between them and therefore does not induce a shift in $\mu_0$.

In the second case (**Figure S5(b)**), the impurity is in direct contact with the graphene. In this configuration, a work-function mismatch at the graphene-impurity interface can modify the local Fermi level of the graphene at that site, which in turn is likely to shift $\mu_0$. Since this impurity also acts as an external dopant, a shift in $n_0$ may accompany the change in $\mu_0$, although the relative magnitude of the two contributions depends on the microscopic details.

Taken together, these considerations suggest that impurities confined at interfaces remote from the graphene manifest as shifts in $n_0$, whereas those adjacent to graphene can additionally shift $\mu_0$. This provides a basis for using spatial maps of $n_0$ and $\mu_0$ to infer the vertical location of electrostatic perturbations such as trapped bubbles.